\documentclass{article}
\usepackage{spconf,amsmath,graphicx,hyperref}
\usepackage{booktabs}
\usepackage{amssymb}

\title{HergNet: a Fast Neural Surrogate Model for Sound Field Predictions via Superposition of Plane Waves}
\name{Matteo Calafà, Yuanxin Xia, Cheol-Ho Jeong}
\address{Acoustic Technology, Department of Electrical and Photonics Engineering, \\ Technical University of Denmark, 2800 Kongens Lyngby, Denmark}

\begin{document}
\maketitle

\begin{abstract}
We present a novel neural network architecture for the efficient prediction of sound fields in two and three dimensions. The network is designed to automatically satisfy the Helmholtz equation, ensuring that the outputs are physically valid. Therefore, the method can effectively learn solutions to boundary-value problems in various wave phenomena, such as acoustics, optics, and electromagnetism. Numerical experiments show that the proposed strategy can potentially outperform state-of-the-art methods in room acoustics simulation, in particular in the range of mid to high frequencies.
\end{abstract}


\section{Introduction}
\label{sec:intro}

Several physical phenomena are represented by propagation of waves, especially in fields like acoustics, optics, quantum mechanics, electromagnetism and surface fluid mechanics \cite{jacobsen2013fundamentals, morse1986theoretical, DeanDalrymple1991, hawkes1996principles, voon2004helmholtz}. Fast and accurate simulations of waves dynamics is therefore of great relevance to the scientific community, in particular in complex scenarios, where high frequencies, broad domains or long time intervals are considered. 

Sound field simulations are commonly carried out by solving the underlying partial differential equation (PDE) models through several available numerical strategies. Specifically, these include the finite element method (FEM) \cite{CRAGGS1994568}, spectral element method (SEM) \cite{pind2019time}, discontinuous Galerkin (DG-FEM) \cite{pind2020time}, boundary element method (BEM) \cite{henriquez2018three} and finite-difference time-domain (FDTD) \cite{kowalczyk2010room}. However, these methods often entail significant computational costs, particularly at high frequencies. Alternative approaches such as geometrical acoustics (GA) \cite{savioja2015overview} approximate sound propagation using rays. While efficient, GA is primarily applicable to high-frequency scenarios and large environments.

In the last years, scientific machine learning (SciML) techniques have emerged as alternatives to traditional numerical solvers \cite{bianco2019machine}. Notable examples include physics-informed neural networks (PINNs) \cite{raissi2019physics,borrel2021physics} and deep neural operators \cite{borrel2024sound}. Although these methods have shown success in tasks such as parameter estimation and repeated parametric simulations, their effectiveness remains limited for direct model-based numerical computations, so that traditional numerical solvers often remain the preferred choice. In response, recent studies have promoted an alternative strategy in which neural networks (NNs) are designed to inherently represent the underlying physics, rather than learning from it \cite{calafa2024physics,calafa2025holomorphic,richter2022neural}. Such hard constraints make possible for NNs to learn only from the boundary conditions, significantly improving the accuracy and the efficiency with respect to classical PINNs.

Building on this idea, we present a novel NN architecture that automatically produces solutions to the wave equation. The introduced strategy applies to both two and three spatial dimensions, and can be applied to several physical problems. In the following, the method is described and tested in learning complex sound fields, demonstrating superior accuracy and performance.

\section{Method}
Let us consider a bounded and closed domain $\Omega \in \mathbb{R}^D$, with $D=2,3$, representing the enclosed space. Under the time-harmonic assumption, we can decompose the acoustic signal into each frequency and rewrite the wave equation as the Helmholtz problem

\begin{equation}\label{eq:helmholtz}
    \nabla^2 p + k^2 p = g, \hspace{5mm} \text{in } \Omega,
\end{equation}

where $p$ is the acoustic pressure, $k=2\pi f/c$ is the wave number associated to the frequency $f$ and sound speed $c$. $g$ denotes instead the sound source, and the velocity field in the direction $\mathbf{n}$ can be obtained by $v_{\mathbf{n}}=i (\nabla p \cdot \mathbf{n}) / (\rho c k)$, where $\rho$ is the mass density of the medium. Sound field calculation typically includes the solving of Equation \eqref{eq:helmholtz} coupled with Robin-type impedance boundary conditions $p = Z v_\mathbf{n}$, where $Z$ is the prescribed locally-reacting acoustic surface impedance on the boundaries and $\mathbf{n}$ is chosen as the outward unit normal vector.

\begin{figure*}[!ht]
\centering
\includegraphics[trim={0 21cm 0 0},clip,width=0.8\linewidth]{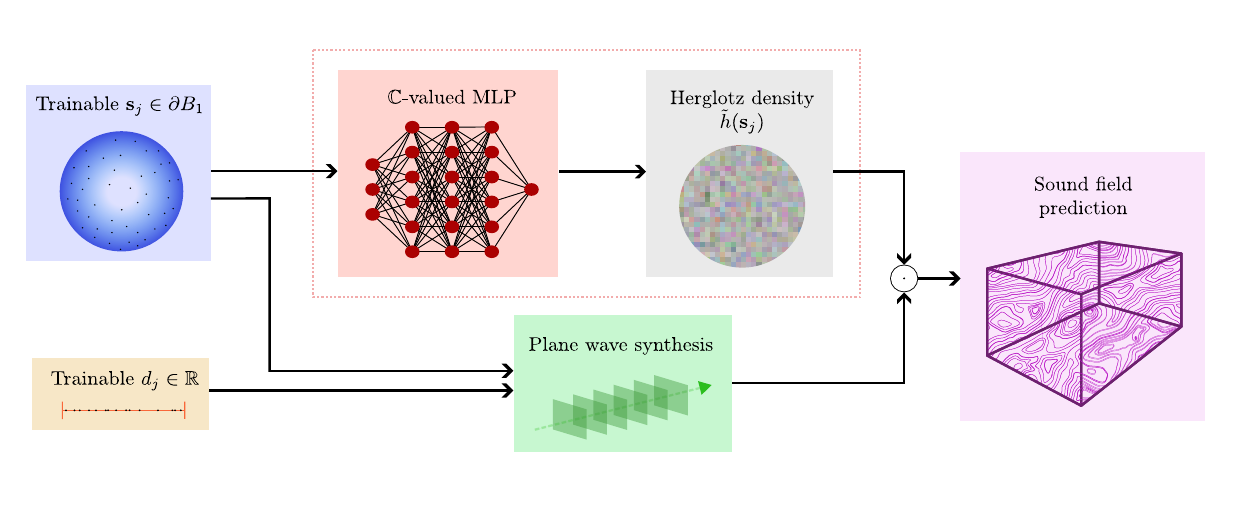}
\caption{Sketch of the HergNet underlying functioning.}
\label{fig:net}
\end{figure*}

$p$ is said to be a \emph{Herglotz} wave function if and only if it can be written as
\begin{equation}\label{eq:herglotz}
    p(\mathbf{x}) = \int_{\partial B_1} e^{i k ( \mathbf{x}\cdot \mathbf{s})} h(\mathbf{s}) \, d\mathbf{s}, \hspace{5mm} \mathbf{x}\in\Omega,
\end{equation}
where $\partial B_1$ is the surface of the unit sphere $B_1=\{\mathbf{x} \in \mathbb{R}^D: |\mathbf{x}|<1\}$ and $h\in L^2(\partial B_1)$ is often called Herglotz density. It is well-known that the space of Herglotz wave functions is dense in the space of the solutions to the homogeneous Helmholtz problem with respect to the $H^1$ norm \cite{colton2001denseness}. In simpler words, regular solutions to the Helmholtz problem in Equation \eqref{eq:helmholtz} with $g=0$ are (possibly infinite) superpositions of plane waves. Vice versa, any superposition of plane waves is a solution to Equation \eqref{eq:helmholtz} by linearity. This property is commonly referred to as plane wave decomposition (or expansion) and has been extensively used in prior studies \cite{perrey2006plane,bianchi2016model,schmid2021spatial}.

The representation in Equation \eqref{eq:herglotz} can be translated into a machine learning framework. Namely, we consider the quantity 
\begin{equation}\label{eq:net}
    \tilde{p}(\mathbf{x}) = \frac{1}{N_{quad}}\sum_{j=1}^{N_{quad}} e^{ik (\mathbf{x}\cdot \mathbf{s}_j + d_j)} \tilde{h}(\mathbf{s}_j), \hspace{5mm} \mathbf{x}\in\Omega,
\end{equation}
where $N_{quad}\in \mathbb{N}$ is a large number, $\mathbf{s}_j\in \partial B_1,d_j\in\mathbb{R}$ are trainable parameters and $\tilde{h}:\partial B_1\rightarrow \mathbb{C}$ is learned through a feedforward neural network. We refer to this framework as HergNet, with a graphical representation shown in Figure \ref{fig:net}.

It is important to note that $\tilde{p}$ can approximate any solution $p$ with respect to the $L^2$ norm, simply by the quadrature approximation of the integral, the universal approximation theorem of NNs \cite{cybenko1989approximation}, and the density of Herglotz functions. Other important observations are worthy of note: first, the parameter $d_j$ is not strictly necessary in the mathematical representation, but it is helpful to achieve better and faster convergence. In addition, the quantity $\tilde{h}(s_j)$ could legitimately be replaced by a scalar trainable parameter, making the approximation problem a simple regression task. However, the application of a neural network to learn the functional relationship $\tilde{h}$ turns out to be necessary, as it embodies the continuity condition between close points $\mathbf{s}_j$, whereas the simpler regression alternative often fails to converge.

In order to account for non-zero source terms $g$, the representation in Equation \eqref{eq:net} is corrected to $\tilde{p}+p_g$, where $p_g$ is any solution to the non-homogeneous problem. For instance, if $g(\mathbf{x})=- \delta(\mathbf{x}-\mathbf{x}_0)$ is a point source, then one can choose $p_g(\mathbf{x}) = G(\mathbf{x}|\mathbf{x}_0)$, being $G$ the monopole fundamental solution.

We define the loss function as the mean squared error (MSE) of the acoustic pressure restricted to the boundary
\begin{equation}\label{eq:loss}
    \mathcal{L}(p) := \mathbb{E}_{\mathbf{x}\sim \mathcal{U}(\partial\Omega)} \left[\left( p (\mathbf{x})-Z(\mathbf{x})v_{\mathbf{n_x}}(\mathbf{x})\right)^2\right],
\end{equation}
where $\mathbf{n_x}$ denotes the outward normal vector at the boundary point $\mathbf{x}$. It is important to emphasize that the loss function in Equation \eqref{eq:loss} contains only boundary terms, in contrast with classical PINNs where internal points are also evaluated, significantly increasing the computational cost.

The parameters $d_j$ are initially uniformly sampled on the interval $[-1,1]$. Similarly, $\mathbf{s}_j$ is sampled uniformly on $\partial B_1$. This is performed for $D=2$ by $\theta_j \sim \mathcal{U}(0,2\pi)$ and $\mathbf{s}_{j}=[\cos(\theta_j),\sin(\theta_j)]$. For $D=3$, we employ the spherical coordinates $\phi_j \sim \mathcal{U}(0,2\pi)$, $\cos(\theta_j) \sim \mathcal{U}(-1,1)$, so that $\mathbf{s}_{j}=[\sin(\theta_j)\cos(\phi_j), \sin(\theta_j)\sin(\phi_j),\cos(\theta_j)]$. $\tilde{h}$ approximates a complex quantity, therefore it is constructed as a complex-valued feedforward fully-connected neural network (CVNN) \cite{hirose2012complex}. Differently from classical PINNs, $\tilde{h}$ needs not to be differentiable, thus the ReLU activation function ReLU$(x)=\max\{0,x\}$ is chosen. Then, the He initialization method is used, i.e.,
\begin{equation*}
    \text{Re}(W),\text{Im}(W) \overset{i.i.d.}{\sim} \mathcal{N}\left(\mathbf{0}, \frac{1}{M}\right),
\end{equation*}
where $W$ is the weight tensor and $M\in\mathbb{N}$ is the number of neurons in the corresponding input layer. We note that the variance is halved with respect to classical real-valued networks \cite{hirose2012complex}.

\section{Results}

The code has been implemented in Python 3.12 by using JAX 0.7.0 \cite{jax2018github}, in particular the NNX Flax API. Tests have been executed on a single NVIDIA GeForce RTX 5090, with 32 GB memory and CUDA 12.8 installed.

We consider a shoebox room with axial lengths according to the Louden's ratio \cite{louden1971dimension} $[0,1]\times [0,1.4]\times[0,1.9]$ m, and a point source located at $[0.2,0.4,0.3]$ m. The frequency-independent surface impedance of the walls is uniform and equal to $Z=(10-10i)\rho c$ Pa$\cdot$s/m, where we assign $c=343$ m/s the speed of sound in air and $\rho = 1.2$ kg/m$^3$ the density of air. Such value yields an absorbing coefficient $\alpha \approx 0.18$, corresponding to a lightly absorbing material.

The neural network used to learn $\tilde{h}$ has two complex-valued inner layers with 10 neurons each. The Adam optimizer \cite{kingma2014adam} is adopted with initial learning rate $2\times 10^{-3}$. The number of training points $N_{train}$ is selected in order to achieve the recommended choice of 6 points per wavelength (PPW). Due to memory limitations, a single batch is used if $N_{train}<5\times 10^4$, two batches are employed otherwise. $N_{quad}$ is also frequency-dependent, and is selected as $N_{quad} = f^2/2000$. It should be noted that both $N_{train}$ and $N_{quad}$ consistently scale as $f^2$, which is expected from the theory and provides a substantial benefit over volume-based methods such as PINNs and FEM. The training is carried out over 1000 epochs.

The analytic Green's function is used as a reference for comparison with the HergNet simulation, since it can be expressed as a superposition of room modes \cite{morse1986theoretical}. However, explicit expressions are available only for perfectly reflecting boundaries. To incorporate realistic surface absorption, a Newton root-finding method is employed to adjust the modes accordingly. Moreover, all modes are included in the series expansion up to twice the problem frequency.

\begin{figure}[!ht]
\centering
\begin{minipage}[b]{0.43\linewidth}
  \centering
  \centerline{\includegraphics[trim={4cm 0 6cm 0},clip,width=\linewidth]{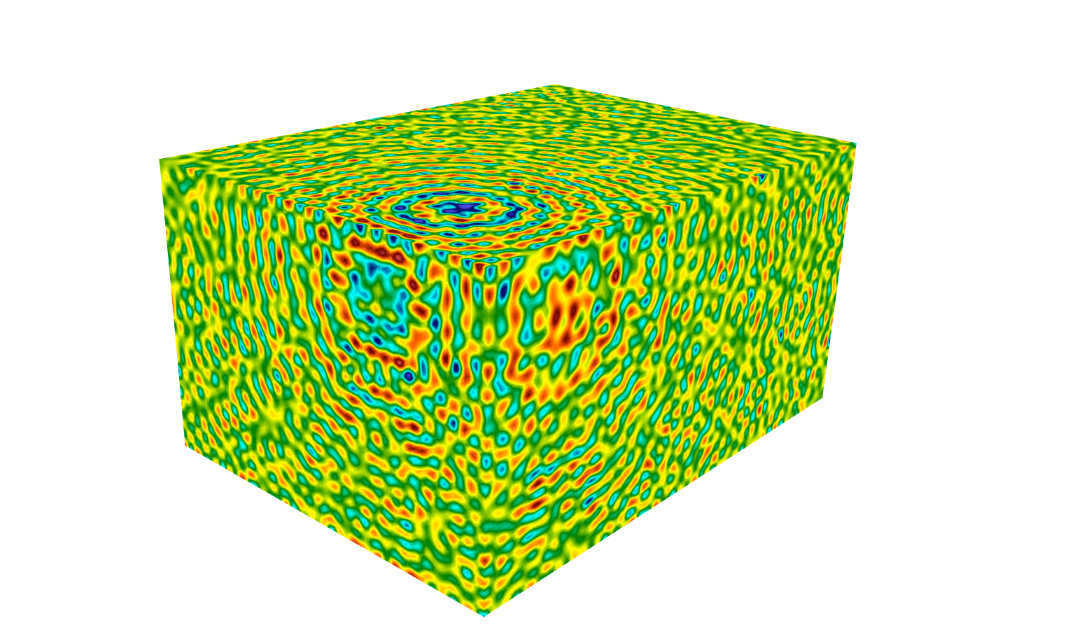}}
  \centerline{(a)}\medskip
\end{minipage}
\hfill
\begin{minipage}[b]{0.56\linewidth}
  \centering
  \centerline{\includegraphics[trim={4cm 0 0 0},clip,width=\linewidth]{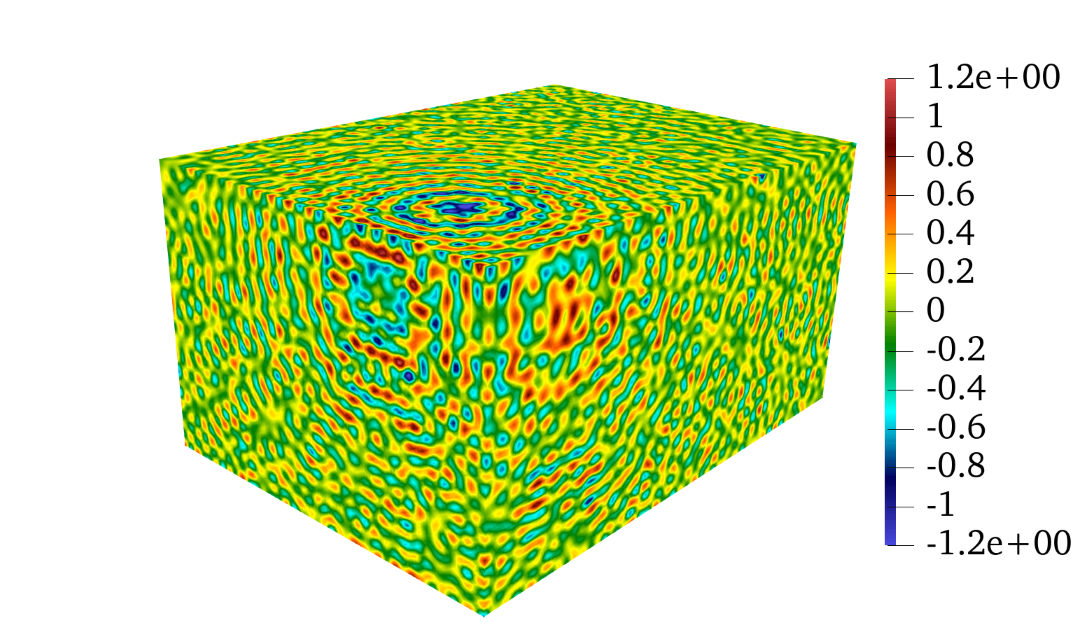}}
  \centerline{(b)}\medskip
\end{minipage}
\\
\begin{minipage}[b]{.43\linewidth}
  \centering
  \centerline{\includegraphics[trim={4cm 0 6cm 0},clip,width=\linewidth]{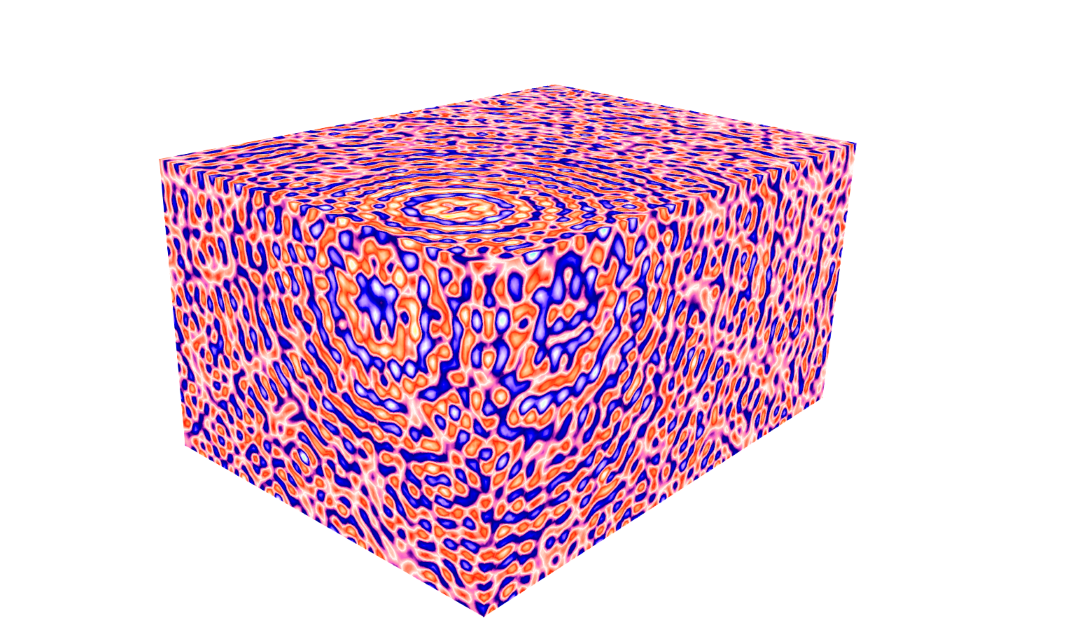}}
  \centerline{(c)}\medskip
\end{minipage}
\hfill
\begin{minipage}[b]{0.56\linewidth}
  \centering
  \centerline{\includegraphics[trim={4cm 0 0 0},clip,width=\linewidth]{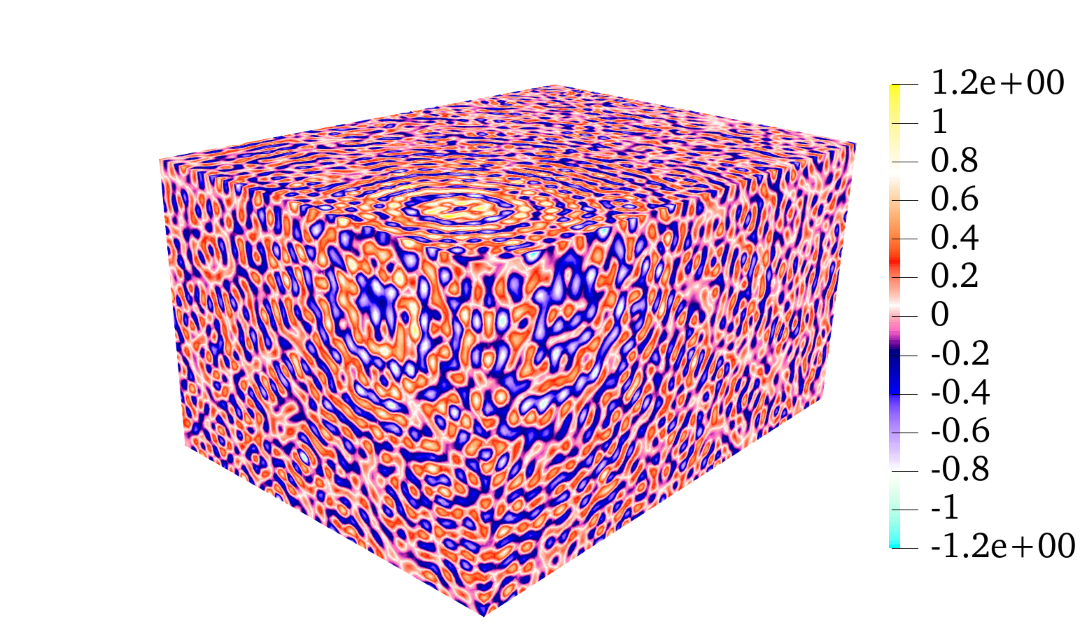}}
  \centerline{(d)}\medskip
\end{minipage}
\\
\centering
\begin{minipage}[b]{0.56\linewidth}
  \centering
  \centerline{\includegraphics[trim={4cm 0 0 0},clip,width=\linewidth]{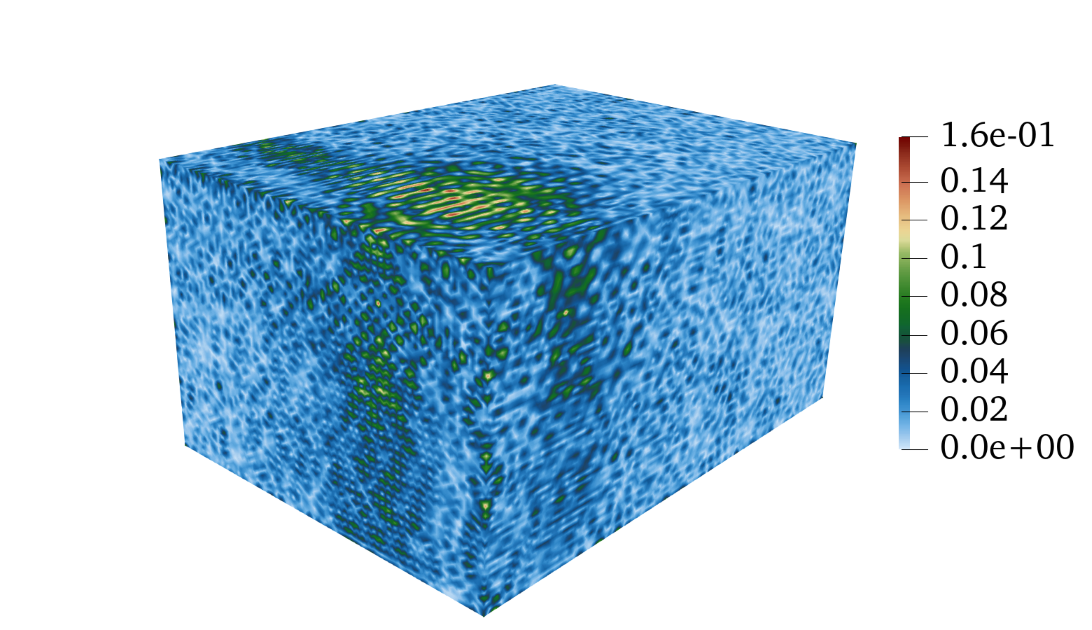}}
  \centerline{(e)}\medskip
\end{minipage}
\caption{Contour plots of the acoustic field at $6000$ Hz (in Pa). (a) Predicted $\text{Re}(p)$ from HergNet. (b) Analytic $\text{Re}(p)$. (c) Predicted $\text{Im}(p)$ from HergNet. (d) Analytic $\text{Im}(p)$. (e) Absolute error.}
\label{fig:contour}
\end{figure}

Figure \ref{fig:contour} shows a qualitative comparison between the simulated sound field and the analytic reference solution for $f=6000$ Hz. In particular, the plots closely match for both the real and imaginary part of the acoustic pressure. The point-wise error is computed as the complex modulus of the difference, yielding a maximum value of $0.16$ Pa on the boundary of the room and $10\%$ maximum relative error. It is important to note that the small discrepancy can be attributed to both terms, as the analytic solution also includes some approximations due to the Netwon's method and the series truncation.

\begin{table}[!ht]
    \centering
    \begin{tabular}{c c c c c}
        \toprule
         Time (s) & Memory (GB) &$N_{quad}$ & $N_{param}$ &  $N_{train}$\\
         \midrule \midrule
          124 & 24.07 & 18000 & 54322 & 131308\\
         \bottomrule
    \end{tabular}
    \vspace{-6pt}
    \caption{Computational cost for the HergNet room acoustics simulation at $6000$ Hz.}
    \label{tab:efficiency}
\end{table}

More details on the computational cost are provided in Table \ref{tab:efficiency}.
The training time is notably low despite the high complexity of the problem, due to the quick evaluation of the forward and backward passes as well as the reduced number of epochs. In contrast, the memory consumption represents the main bottleneck. In particular, in the worst single-batch scenario, all training points are expanded in the stacked $N_{quad}$ dimension, leading to a fast-growing $\mathcal{O}(f^4)$ required memory space. The total number of trainable parameters $N_{param}$ scales with $N_{quad}$, entailing a moderately large quantity.

\begin{figure*}[!ht]

\begin{minipage}[b]{0.49\linewidth}
  \centering
  \centerline{\includegraphics[trim={0 0 0 0},clip,width=\linewidth]{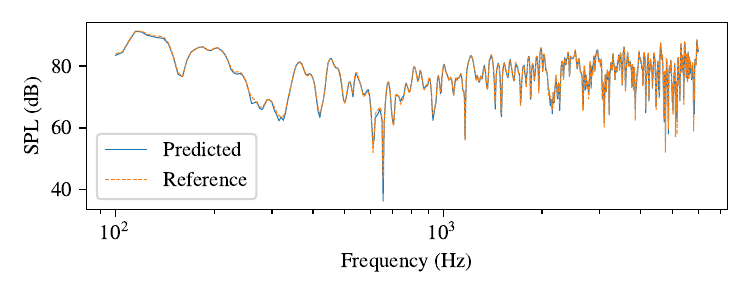}}
   \centerline{(a)}
\end{minipage}
\hfill
\begin{minipage}[b]{0.49\linewidth}
  \centering
  \centerline{\includegraphics[width=\linewidth]{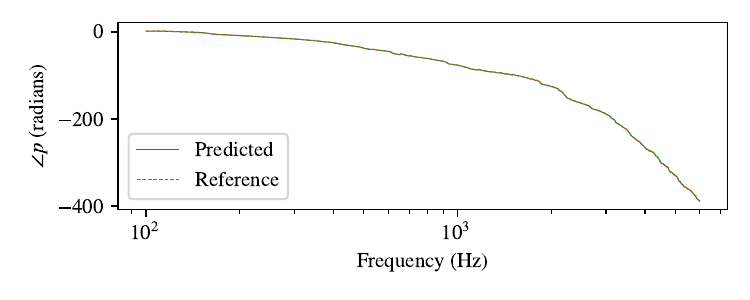}}
     \centerline{(b)}
\end{minipage}

\begin{minipage}[b]{0.49\linewidth}
  \centering
  \centerline{\includegraphics[width=\linewidth]{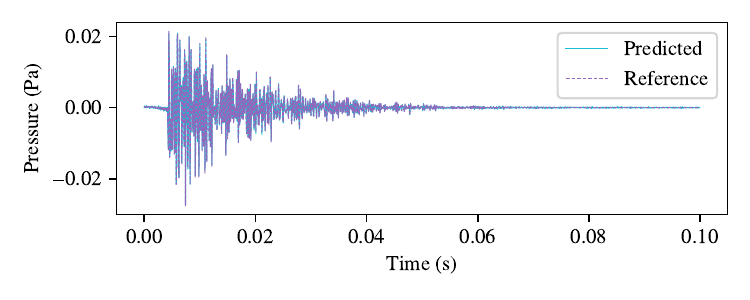}}
     \centerline{(c)} 
\end{minipage}
\hfill
\begin{minipage}[b]{0.49\linewidth}
  \centering
  \centerline{\includegraphics[width=\linewidth]{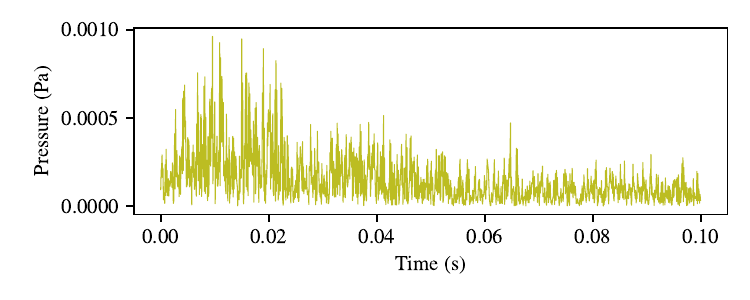}}
       \centerline{(d)}
\end{minipage}
\caption{Acoustic signal for source at $[0.2,0.4,0.3]$ m and receiver at $[0.7, 1.2, 1.5]$ m. Transfer function in terms of (a) SPL and (b) phase. (c) Impulse response. (d) Impulse response error.}
\label{fig:signal}
\end{figure*}

On the other hand, these results mark a substantial improvement over computational methods that require evaluations throughout the interior volume of the room. For example, a FEM simulation using the recommended 6 elements per wavelength would necessitate over 3 million elements, resulting in a relatively high cost. Similarly, classical PINNs would demand substantially greater memory, longer training times, and are unlikely to converge without the use of advanced techniques. Previous studies have indeed addressed simpler tests, with reported training times ranging from several minutes to even hours \cite{borrel2021physics,schmid2023physics}. These considerations highlight the rationale for preferring the analytic solution to FEM solutions, when comparing HergNet solutions.

A frequency sweep is performed from $100$ Hz up to $6000$ Hz by applying the same configurations listed above. The frequency resolution is $5$ Hz, given the estimated reverberation time of $T_{60} \approx 0.2$ s. For lower frequencies, a minimum value of 1000 is assigned to $N_{quad}$ and $N_{train}$. The sound field evaluation is performed at the coordinates $[0.7, 1.2, 1.5]$ m, thus approximately at the opposite side with respect to the point source. 

Figure \ref{fig:signal} shows the comparison of the simulated and analytic solution. A perfect match is notable in the unwrapped phase in the entire spectrum. The sound pressure level (SPL), computed as $20\log_{10}(|p|/|p_{ref}|)$, being $p_{ref}=2\times 10^{-5}$ Pa, also aligns closely, particularly in the mid- to high-frequency range. Deviations remain well below the just-noticeable difference (JND) threshold of 1 dB, except in correspondence of the dips. These regions are indeed known to be especially sensitive to numerical errors in acoustic simulations \cite{lam2005issues}. However, because such discrepancies are less perceptible to human hearing, the HergNet results remain perceptually identical to the analytic solution. The impulse response was also obtained by using the inverse fast Fourier transform (i-FFT), demonstrating an excellent agreement. Specifically, relative errors are largely below $10\%$.

It is worth noting that lower frequencies are associated with slightly larger errors, as evidenced by the more irregular numerical profiles compared to the smoother analytic solution. This behavior is in contrast with the reduced complexity of the problem at small wave numbers, where fewer training points and parameters are required. The increased error is therefore likely related to the underlying mathematical representation in connection to the physical behavior of the problem. Specifically, at low frequencies the PDE behaves predominantly as a diffusion problem, with the limiting case $k\rightarrow0$ reducing to the pure Laplace equation. In contrast, the plane wave superposition method enforces the oscillatory behavior, making it more suitable for higher frequency regimes. This limitation is further supported by inspecting Equation \eqref{eq:herglotz}, which shows that non-constant low-frequency sound fields necessarily correspond to diverging Herglotz densities.

\section{Conclusions}
The proposed plane wave superposition learning approach, grounded in the Herglotz representation framework, has proven to be both effective and accurate in simulating sound fields. Although tests are restricted to acoustics, the method is easily generalizable to different scientific applications, in both two and three spatial dimensions. Notably, results exhibit strong stability and fast convergence, even in challenging high-frequency scenarios, outperforming classical PINNs and achieving comparable performance with state-of-the-art numerical algorithms. More specifically, the approach is particularly well-suited for mid- to high-frequency regimes, while its effectiveness is slightly reduced at lower frequencies. Future work may focus on extending the method performance across the entire spectrum, although it is worth noting that several existing wave-based numerical approaches already achieve high efficiency in the low-frequency regime.

\clearpage
\newpage
\ninept

\bibliographystyle{IEEEbib}
\bibliography{bib}

\end{document}